\documentclass[authoryear,12pt,letter,3p]{jowarticle}
\usepackage{graphicx}
\usepackage{amsmath}
\usepackage{amssymb}
\usepackage{setspace}
\usepackage[all]{xy}
\usepackage{amsthm}
\usepackage{appendix}
\usepackage{csquotes}
\usepackage{longtable}
\usepackage{tabu}
\makeatletter
\renewcommand\section{\@startsection{section}{1}{\z@}{-3.25ex plus -1ex minus -.2ex}{1.5ex plus .2ex}{\normalsize\bf}}
\renewcommand\subsection{\@startsection{subsection}{2}{\z@}{-3.25ex plus -1ex minus -.2ex}{1.5ex plus .2ex}{\normalsize\bf}}
\renewcommand\subsubsection{\@startsection{subsubsection}{3}{\z@}{-3.25ex plus -1ex minus -.2ex}{1.5ex plus .2ex}{\normalsize\bf}}
\makeatother

\newtheorem{thm}{Theorem}

\newtheorem{prop}[thm]{Proposition}

\begin{document}

\begin{frontmatter}
\title{Maxwell-Huygens, Newton-Cartan, and Saunders-Knox Spacetimes}
\author{James Owen Weatherall}\ead{weatherj@uci.edu}
\address{Department of Logic and Philosophy of Science\\ University of California, Irvine, CA 92697}
\begin{abstract} I address a question recently raised by Simon Saunders [\emph{Phil. Sci.} \textbf{80}(2): 22-48 (2013)] concerning the relationship between the spacetime structure of Newton-Cartan theory and that of what I will call ``Maxwell-Huygens spacetime''.  This discussion will also clarify a connection between Saunders' work and a recent paper by Eleanor Knox [\emph{Brit. J. Phil. Sci.} \textbf{65}(4): 863-880 (2014)].
\end{abstract}
\begin{keyword}Newton-Cartan theory; classical spacetime; Maxwellian spacetime; Maxwell-Huygens spacetime; Newtonian gravitation
\end{keyword}
\end{frontmatter}

\doublespacing

In a recent paper, Simon \citet{Saunders} argues that proper reflection on Corollary VI to Newton's laws of motion reveals that the spacetime structure presupposed by Newton's laws is one on which there is a privileged class of ``non-rotating'' states of motion, but \emph{not} a privileged class of non-accelerating, or inertial, states of motion.  Thus Saunders' view differs markedly from the received view, according to which Newton's laws presuppose Galilean spacetime, a structure in which one does have a privileged class of inertial trajectories \citep{Stein, Earman, DiSalle}.  Saunders calls his alternative structure ``Newton-Huygens spacetime'', though he acknowledges that the same structure was introduced by \citet[\S 2.3]{Earman} under the moniker ``Maxwellian spacetime''.\footnote{Actually, Saunders uses ``Newton-Huygens spacetime'' to refer to the spacetime structure understood as a modified affine space; Maxwellian spacetime refers to its characterization in terms of fields on a manifold.  For present purposes, the difference is inconsequential and I will limit attention to the latter.} (As a compromise, I will call it ``Maxwell-Huygens spacetime'' here.)

At the end of his paper, Saunders poses the following question: ``What is the relation between a theory of gravity (and other forces) formulated in [Maxwell-Huygens] space-time and one based on Newton-Cartan space-time?'' \citep[p. 46]{Saunders}.  It is this question that I hope to answer in what follows.  Doing so, I will argue, reveals a certain inadequacy in Saunders' account of classical spacetime structure.  It also shows a close, and perhaps not altogether obvious, connection between Saunders' work and recent work by \citet{Knox}.

To begin, I will give a precise characterization of Maxwell-Huygens spacetime.  This structure, which may be denoted by $(M,t_a,h^{ab},[\nabla])$, consists in a smooth four dimensional manifold $M$, which we take to be diffeomorphic to $\mathbb{R}^4$; a ``temporal metric'' $t_{a}$, which is a smooth one form of signature $(1,0,0,0)$;\footnote{By defining the temporal metric as a one form, we have implicitly assumed that Maxwell-Huygens spacetime is ``temporally oriented''.  See \citet[p. 251]{MalamentGR}.  Note, too, that here (and throughout) we are working in the abstract index notation, which is explained in \citet[\S1.4]{MalamentGR}.} a spatial metric $h^{ab}$, which is a smooth symmetric tensor field of signature $(0,1,1,1)$ satisfying $h^{ab}t_{a}=\mathbf{0}$; and a ``standard of rotation,'' which I will denote by $[\nabla]$.  The temporal and spatial metrics allow one to distinguish between ``timelike vectors'' (vectors $\xi^a$ at a point $p\in M$ for which $\xi^at_a\neq 0$) and ``spacelike vectors'' (any other vector); to assign temporal lengths to arbitrary vectors at a point; and to assign spatial lengths to spacelike vectors at a point.  This structure also permits one to foliate spacetime into constant time hypersurfaces; in what follows, we assume these hypersurfaces are diffeomorphic to $\mathbb{R}^3$ and complete relative to the metric induced by $h^{ab}$.

The temporal and spatial metrics are now standard notions. (See \citet[\S 4.1]{MalamentGR} or \citet[Ch. 2]{Earman} for further details.)  The standard of rotation, however, requires further comment.  The idea is that we want enough structure to say whether a smooth timelike vector field $\xi^a$, representing, say, the motion through spacetime of bits of some body, is ``rotating''.  The class of non-rotating vector fields, then, would represent the privileged states of non-rotation mentioned above.

To capture the idea of a standard of rotation, let me begin by describing Galilean spacetime in more detail \citep[\S2.4]{Earman}.\footnote{See also \citet[\S 4.1]{MalamentGR}.  For Malament, the structures I call Galilean spacetimes are ``classical spacetimes'' with a flat derivative operator.}  Galilean spacetime consists in a structure that may be written $(M,t_a,h^{ab},\nabla)$, where $M$, $t_a$, and $h^{ab}$ are all as described above.  The additional piece of structure is a flat covariant derivative operator $\nabla$ on $M$ that we require to be compatible with the spatial and temporal metrics in the sense that $\nabla_a t_b=\mathbf{0}$ and $\nabla_a h^{bc}=\mathbf{0}$.\footnote{Here and throughout we restrict attention to torsion-free derivative operators.}  Such an object provides a standard of constancy for vector fields; it also provides a notion of inertial motion, corresponding to the constant timelike vector fields.  By providing a standard of non-acceleration, a derivative operator also provides a natural standard of rotation.  This can be captured as followed.  Given a unit timelike vector field $\xi^a$ (i.e., a field such that $|\xi^at_a|=1$) in Galilean spacetime, we may then say that $\xi^a$ is \emph{non-rotating} if and only if $\nabla^{[a}\xi^{b]}=\mathbf{0}$, where $\nabla^a\xi^b=h^{an}\nabla_n\xi^b$.\footnote{The brackets indicate anti-symmetrization over the indices. More generally, we may define a \emph{rotation tensor} $\omega_{ab}$ associated with $\xi^a$ that captures the magnitude and direction of its rotation; this tensor vanishes iff the non-rotation condition just stated holds. See \citet[p. 264]{MalamentGR} for details.}

Defining non-rotation in this way certainly allows us to make the determinations we require.  But in fact it provides \emph{too much} structure, since as we have seen, a derivative operator also allows one to define a class of preferred inertial states of motion.  What we wanted was a standard of rotation and no more.  One way of making precise the sense in which a derivative operator gives us strictly more structure than just a standard of rotation is to observe that in general \emph{many} flat derivative operators are compatible with the classical metric structure,\footnote{Note that this is a strong disanalogy between the metrical structure of classical spacetime and that of (pseudo-)Riemannian geometry.  See \citet[Prop. 4.1.3]{MalamentGR}.} and moreover, many of these will agree on whether a given vector field is non-rotating, even though they disagree on other matters, such as whether a given vector field is constant.  This suggests that we can capture the idea of a standard of rotation by considering an equivalence class of flat, metric-compatible derivative operators, all of which agree on whether any given vector field is rotating: \[[\nabla]=\{\nabla':R'^a{}_{bcd}=\mathbf{0};\nabla'_at_b=\mathbf{0};\nabla'_ah^{bc}=\mathbf{0}\text{; and }|\xi^at_a|=1 \Rightarrow(\nabla'^{[a}\xi^{b]}=\mathbf{0}\Leftrightarrow\nabla^{[a}\xi^{b]}=\mathbf{0})\}.\footnote{Here $R'^a{}_{bcd}$ is the Riemann curvature tensor associated with $\nabla'$; requiring this tensor to vanish makes precise the requirement that $\nabla'$ is flat.  For more on curvature tensors, see \citet[\S1.8]{MalamentGR}.} \]
Indeed, this is how Earman characterizes Maxwellian spacetime as well.


Let us now turn to Newton-Cartan theory, which is a theory both of spacetime structure and of gravitation.\footnote{For background on Newton-Cartan theory, see \citet{Trautman} and \citet[Ch. 4]{MalamentGR}.}  It will once again be helpful to begin with Galilean spacetime.  In that setting, one may think of Newtonian gravitation as a theory relating a gravitational potential, which is represented by a scalar field $\varphi$ on $M$, and the distribution of mass density in spacetime, represented by a scalar field $\rho$.  These together satisfy Poisson's equation, which may be written as $\nabla_a\nabla^a\varphi=4\pi\rho$.  In the presence of (only) a gravitational potential, massive point particles will traverse trajectories with acceleration given by $\xi^n\nabla_n\xi^a = -\nabla^a\varphi$, where $\xi^a$ is the tangent to the particle's worldline.  Models of the theory may be written $(M,t_a,h^{ab},\nabla,\varphi)$.\footnote{Here and throughout, I am suppressing $\rho$ in specifications of models of both Newton-Cartan theory and Newtonian gravitation, since one can uniquely recover $\rho$ from the other fields and the geometrized Poisson and Poisson equations, respectively.}

Models of Newton-Cartan theory, meanwhile, which I will call ``Newton-Cartan spacetimes,'' may be written as $(M,t_a,h^{ab},\tilde{\nabla})$, where $M$, $t_a$, and $h^{ab}$ are as above, and $\tilde{\nabla}$ is a derivative operator compatible with $t_a$ and $h^{ab}$.  Unlike in Galilean spacetimes, $\tilde{\nabla}$ is no longer required to be flat.  Instead, its associated Ricci curvature tensor is required to satisfy $\tilde{R}_{ab}=4\pi\rho t_a t_b$, where $\rho$ is again some smooth scalar field representing the mass density.\footnote{For a given derivative operator $\nabla$, the associated Ricci tensor is defined by $R_{ab}=R^n{}_{abn}$, where $R^{a}{}_{bcd}$ is the Riemann curvature tensor.}  In this theory, as in general relativity, gravitation is not conceived as a force; instead, it is a manifestation of spacetime curvature, in the sense that, in the absence of external forces, bodies will traverse timelike geodesics of the curved derivative operator.  It is common to assume that $\tilde{\nabla}$ satisfies two further curvature conditions, which we will take for granted in what follows: $\tilde{R}^a{}_b{}^c{}_d=\tilde{R}^c{}_d{}^a{}_b$ and $\tilde{R}^{ab}{}_{cd}=\mathbf{0}$.\footnote{These two conditions are necessary for the Trautman recovery theorem below.  For more on the significance of the conditions, see \citet[\S 4.3]{MalamentGR}; for the consequences of dropping the second condition, see \citet[\S 4.5]{MalamentGR}.}  Note that although a model of Newton-Cartan theory amounts to a spacetime structure in much the same way that Maxwell-Huygens spacetime and Galilean spacetime do, it also does a bit more, since it determines the allowed trajectories of bodies \emph{even in the presence of other matter}.  In Galilean spacetime, one needs an additional field---the gravitational potential---to determine those allowed trajectories.

Newtonian gravitation in Galilean spacetime and Newton-Cartan theory are systematically related.  The relationship is given by the following two results, originally due to \citet{Trautman}.
\begin{thm}[Geometrization Lemma]\label{geometrization} Let $(M,t_a,h^{ab},\nabla)$ be a Galilean spacetime and let $\varphi$ and $\rho$ be smooth scalar fields on $M$ satisfying $\nabla_a\nabla^a\varphi=4\pi\rho$.  Then there exists a unique derivative operator $\tilde{\nabla}$ such that (1) $(M,t_a,h^{ab},\tilde{\nabla})$ is a model of Newton-Cartan theory with mass density $\rho$ and (2) for all timelike curves with unit tangent field $\xi^a$, $\xi^n\nabla_n\xi^a=-\nabla^a\varphi$ if and only if $\xi^n\tilde{\nabla}_n\xi^a=\mathbf{0}$.  \end{thm}
\begin{thm}[Recovery Theorem]\label{recovery} Let $(M,t_a,h^{ab},\tilde{\nabla})$ be a model of Newton-Cartan theory with mass density $\rho$. Then there exists a smooth scalar field $\varphi$ and a derivative operator $\nabla$ such that (1) $(M,t_a,h^{ab},\nabla)$ is a Galilean spacetime, (2) $\nabla$ and $\varphi$ satisfy $\nabla_a\nabla^a\varphi=4\pi\rho$, and (3) for all timelike curves with unit tangent field $\xi^a$, $\xi^n\nabla_n\xi^a=-\nabla^a\varphi$ if and only if $\xi^n\tilde{\nabla}_n\xi^a=\mathbf{0}$.  Moreover, the pair $(\nabla,\varphi)$ is not unique.  A second pair $(\nabla',\varphi')$ will satisfy the same conditions iff (1) $\nabla^a\nabla^b(\varphi'-\varphi)=\mathbf{0}$ and (2) $\nabla'=(\nabla,C^a{}_{bc})$, where $C^a{}_{bc}=t_a t_b\nabla^a(\varphi'-\varphi)$.\end{thm}
\noindent Given a model of Newton-Cartan theory, we will will call the associated models of Newtonian gravitation in Galilean spacetime given by Theorem \ref{recovery} ``Trautman recoveries''.

I am now in a position to state a first result concerning the relationship between Newton-Cartan theory and Maxwell-Huygens spacetimes.
\begin{prop}\label{MaxwellRecovery}Let $(M,t_a,h^{ab},\tilde{\nabla})$ be a model of Newton-Cartan theory satisfying the conditions of Theorem \ref{recovery} and let $(M,t_a,h^{ab},\nabla,\varphi)$ be a Trautman recovery of this spacetime.  Then given any other flat derivative operator $\nabla'$ on $M$ that is compatible with $t_a$ and $h^{ab}$, there exists a smooth scalar field $\varphi'$ such that $(M,t_a,h^{ab},\nabla',\varphi')$ is a Trautman recovery of $(M,t_a,h^{ab},\tilde{\nabla})$ iff $\nabla$ and $\nabla'$ agree on a standard of rotation.\end{prop}
\noindent The proof of this proposition appears in an appendix.  The proposition provides a precise characterization of one relationship between Maxwell-Huygens spacetime and Newton-Cartan theory: namely, given a Newton-Cartan spacetime, the collection of flat derivative operators corresponding to Trautman recoveries determines a unique standard of rotation $[\nabla]$.

Of course, as I noted above, \emph{any} derivative operator generates a standard of rotation, in the sense that relative to any derivative operator, one can determine whether a given smooth timelike vector field $\xi^a$ is rotating.  So it is not surprising that the derivative operator associated with Newton-Cartan theory does so.  But as we also noted, a derivative operator provides more than a standard of rotation.  What is striking here, then, is not that a Newton-Cartan spacetime allows one to determine whether a given vector field (or body) is rotating, but rather that the \emph{only} structure agreed upon by the flat derivative operators associated with Trautman recoveries of $(M,t_a,h^{ab},\tilde{\nabla})$ is the standard of rotation $[\nabla]$.\footnote{Of course, they also agree on parallel transport of spacelike vectors, and on other structure that all elements of $[\nabla]$ agree on, such as flatness.}  In other words, although a Newton-Cartan spacetime gives rise to many Galilean spacetimes via the recovery theorem, there is a sense in which it naturally gives rise to a \emph{unique}, privileged Maxwell-Huygens spacetime, insofar as that Maxwell-Huygens spacetime is the shared spacetime structure associated with all of the Trautman recoveries of a given Newton-Cartan spacetime.

It is worth noting an important way in which this result depends on the definition of a standard of rotation above.  Following Earman and Saunders (insofar as he cites Earman's definition), I defined a standard of rotation as an equivalence class of \emph{flat} derivative operators.  But one might imagine different definitions, which might not yield results as clean as Prop. \ref{MaxwellRecovery}.  For instance, if one defined a standard of rotation as an equivalence class of derivative operators with arbitrary curvature (satisfying the other conditions), one would not recover the full equivalence class in the way described by Prop. \ref{MaxwellRecovery}, since all Trautman recoveries involve flat derivative operators.

This last observation raises several interesting questions.  For instance, are there different, essentially inequivalent, ways of characterizing a standard of rotation?\footnote{\citet{MalamentRotation} takes up a related question, of whether there is an unambiguous criterion for when a body is rotating that answers to all of our classical desiderata, in the context of general relativity, and finds that the answer is ``yes''.  Here the issue is somewhat different, since we already have one criterion that apparently \emph{does} answer to all our desiderata---namely, the one described above.  But this criterion is described by considering an equivalence class of derivative operators, and one might wonder if (a) there is an intrinsic characterization of a standard of rotation that does not require us to appeal to structure we then forget, and (b) if so, does it (always) determine a unique equivalence class of derivative operators, as described here.}  And, if we do not limit attention to Trautman recoveries, is there still a sense in which the only derivative operators relative to which the geodesics of a given Newton-Cartan spacetime may be naturally expressed as ``accelerated by some force field'' those that agree with that Newton-Cartan spacetime on a standard of rotation?  I will not address the first question here, but an answer to the second is essentially given by Prop. 1 of \citet{Weatherall+Manchak}.  There it is shown that given a Newton-Cartan spacetime $(M,t_a,h^{ab},\tilde{\nabla})$ and \emph{any} derivative operator $\nabla$ compatible with the classical metrics, there is a smooth, antisymmetric field, $F_{ab}$, naturally interpreted as a ``force field'', such that for any timelike curve with unit tangent field $\xi^a$, $\xi^n\tilde{\nabla}_n\xi^a = \mathbf{0}$ if and only if $\xi^n\nabla_n\xi^a=h^{am}F_{mn}\xi^n$.  In particular, $\nabla$ need \emph{not} agree with $\tilde{\nabla}$ on a standard of rotation.  This helps clarify just what is shown in Prop. \ref{MaxwellRecovery}: we get a unique Maxwell-Huygens spacetime only after we limit attention to flat derivative operators and require that the force field be expressible as in ordinary Newtonian gravitation, that is, as the gradient of a scalar field, rather than as a more general object, such as $F_{ab}$.

What about going in the other direction?  Given a Maxwell-Huygens spacetime, can we always recover a unique Newton-Cartan spacetime?  Not as we have set things up thus far.  But the reason is clear: Newton-Cartan theory is a gravitational theory, whereas Maxwell-Huygens spacetime is a spacetime structure in which a gravitational theory may be expressed.   The question, then, is how to express Newtonian gravitation in Maxwell-Huygens spacetime.  To do so, one will need to provide some prescription for determining how a body will move, given some distribution of masses in the universe.  Certainly our ordinary ways of doing this will no longer work, since spacetime is not endowed with a fixed derivative operator.  In other words, since Maxwell-Huygens spacetime does not have any privileged class of inertial trajectories relative to which one can describe acceleration, one cannot make sense of a distinction between forced and unforced motion.  In particular, then, Newtonian gravitation may not be conceived as a theory of gravitational force in anything like the standard sense.\footnote{Of course, \citet{Saunders} also develops a way of determining how bodies move, by characterizing the evolution of \emph{relative} positions of particles in Maxwell-Huygens spacetime, including in the presence of of forces, such as gravitational force, that depend only on the distances between particles.  Importantly, this approach requires a radical reconceptualization of force as proportional to relative acceleration, not absolute acceleration.  In any case, for present purposes, as long as everyone agrees on what the trajectories that bodies follow \emph{are}, how one arrives at those trajectories is inconsequential, and the strategy pursued below, though less elegant than Saunders', is more perspicuous for the point I aim to make.}

That said, Prop. \ref{MaxwellRecovery} provides a natural indirect option.  That proposition, in conjunction with Theorem \ref{recovery} and standard results concerning the existence of solutions to differential equations, shows that given a Maxwell-Huygens spacetime $(M,t_a,h^{ab},[\nabla])$, for any $\nabla\in[\nabla]$, there exists some scalar field $\varphi$ such that (1) $\nabla_a\nabla^a\varphi=4\pi\rho$, where $\rho$ is the mass density distribution of spacetime,\footnote{Note that nothing in particular depends on $\rho$ being a smooth scalar field---one might just as well define the gravitational potential $\varphi$ as for point particles, using the relation $\varphi=GM/r$, and then work locally.} and (2) the allowed trajectories of bodies are curves $\gamma$ whose acceleration (relative to $\nabla$) is given by $\xi^n\nabla_n\xi^a=\nabla^a\varphi$.  Note that, since Poisson's equation admits homogeneous solutions, there will not be a \emph{unique} scalar field $\varphi$ satisfying Poisson's equation for a given mass density field $\rho$; moreover, different choices may lead to different trajectories---and thus to different geometrizations, in the sense of Theorem \ref{geometrization}.\footnote{Note that this means that we do \emph{not} have a unique recovery of a Newton-Cartan spacetime from a given Maxwell-Huygens spacetime, even after stipulating a mass density $\rho$.  One needs to do a bit more.}  But in this regard, we are in precisely the same situation as in ordinary Newtonian gravitation, and the same sorts of heuristic methods for determining a gravitational potential apply: in some cases, for instance, we might choose to work with a representative from $[\nabla]$ according to which the center of mass of some system of interest follows a constant trajectory, and then impose boundary conditions on $\varphi$ so that it vanishes ``at infinity''.  In the present context, however, such heuristics have merely pragmatic value, providing a way of calculating the allowed trajectories given some distribution of matter.  Similarly, we do not interpret the acceleration of a curve relative to $\nabla$ as representing some fact of the matter about the curve; nor do we need to interpret the gravitational potential or corresponding gravitational field, $\nabla^a\varphi$, as representing facts about force or a field-like entity.

What is the invariant physical structure in this theory?  For one, as we have seen, there is the standard of rotation shared between the derivative operators.  This gives the sense in which this is a theory in Maxwell-Huygens spacetime.  The other invariant structure, however, is the collection of allowed trajectories for bodies.  These are calculated in different ways depending on which representative one chooses from $[\nabla]$, and the accelerations associated with each such curve varies similarly.  So we do not have the structure to say that these curves are accelerating or not.  But however they are described, i.e., whatever acceleration (if any) is attributed to them, the curves themselves are fixed.  Indeed, given some distribution of matter in spacetime, it is these curves that form the empirical content of Newtonian gravitational theory.

Now suppose that we are given some such collection of curves, $\{\gamma\}_{\rho}$, relativized to a matter distribution $\rho$, in Maxwell-Huygens spacetime.  Suppose, too, that however these curves are determined---whether by the calculational procedure just mentioned or some other method---they agree with the possible trajectories allowed by ordinary Newtonian gravitation. It turns out that with \emph{this} information, one \emph{can} uniquely reconstruct a Newton-Cartan spacetime.
\begin{prop}\label{NCrecovery}Let $\{\gamma\}_{\rho}$ be the collection of allowed trajectories for a given mass distribution $\rho$ in Maxwell-Huygen spacetime $(M,t_a,h^{ab},[\nabla])$, as described above.  Then there exists a unique derivative operator $\tilde{\nabla}$ such that (1) $\{\gamma\}_{\rho}$ consists in the timelike geodesics of $\tilde{\nabla}$ and (2) $(M,t_a,h^{ab},\tilde{\nabla})$ is a model of Newton-Cartan theory for mass density $\rho$.\end{prop}
\noindent Again, the proof appears in the appendix.

This last proposition provides an even stronger relationship between a gravitational theory in Maxwell-Huygens spacetime and Newton-Cartan theory.  In a sense, such a gravitational theory simply \emph{is} Newton-Cartan theory, insofar as given the invariant structure of such a theory---namely, the allowed trajectories---one can uniquely construct a derivative operator such that those trajectories are the geodesics of the derivative operator.\footnote{For further argument to the effect that Newton-Cartan theory should be construed as equivalent to Newtonian gravitation once one has accepted that different choices of flat derivative operator---and thus, of inertial frames---may be physically equivalent, see \citet{Weatherall}.  For reasons described here, it is very natural to interpret Maxwell-Huygens spacetime as the spacetime structure presupposed by the theory described there as NG$_2$.}  Note, too, that this result---at least as I interpret it here---reveals a certain inadequacy in Saunders' account.  Saunders insists that there is \emph{no} privileged standard of acceleration in Maxwell-Huygens spacetime.  And there are a few senses in which that is right: (1) before accounting for gravitational influences, Maxwell-Huygens spacetime does not have enough structure to make sense of acceleration; and (2) even in the presence of dynamical considerations, there is in general no privileged \emph{flat} derivative operator, and thus no privileged collection of inertial frames in the standard sense, relative to which acceleration may be defined.  Nonetheless, it turns out that once one takes the dynamically allowed trajectories into account, one \emph{can} define a standard of acceleration, namely, the unique one relative to which the allowed trajectories are geodesics.

In retrospect, this result should not be surprising.  After all, as noted above, there are \emph{empirically} privileged trajectories---namely, those that actual bodies follow, given some mass density distribution.  Whether one is on such a trajectory may be determined by using a simple device, such as an inert mass suspended by springs in the center of a cube-shaped frame.  The privileged trajectories, then, are the ones relative to which the mass stays centered in the frame.  In the context of relativity theory, we call such devices ``accelerometers''; more generally, given how we usually think of acceleration and force in both relativity theory and Newtonian gravitation, it is natural to think of trajectories on which the mass is stationary as unaccelerated, since it is on these trajectories that the springs need not exert any force on the mass.  And indeed, it is precisely these trajectories that are picked out as unaccelerated by the Newton-Cartan derivative operator.  From this point of view, there is always a rough-and-ready way to determine a class of privileged trajectories, even if these are not inertial frames in the traditional sense.

I will conclude by elaborating on the connection with \citet{Knox} mentioned above.  Like Saunders, Knox begins by meditating on the significance of Corollary VI for how we should think of spacetime structure in Newtonian gravitation.  Her conclusion, however, seems quite different from Saunders': she argues that Galilean spacetime is merely a halfway point on the journey from Newtonian spacetime to Newton-Cartan theory.\footnote{Newtonian spacetime is Galilean spacetime plus a privileged standard of rest, given by a constant timelike vector field.  See \citet[\S 2.5]{Earman}.}  This presents a prima facie puzzle: whence the difference?  But I take the arguments here to dissolve the puzzle: Knox and Saunders do not end up in different places after all, except insofar as Knox takes gravitational influences into account all the way through her analysis.  Once one fully considers the effects of gravitation in Maxwell-Huygens spacetime, Newton-Cartan theory is precisely the result.

\appendix

\section{Proofs of propositions}

\noindent\textbf{Proof of Prop. \ref{MaxwellRecovery}}.

We begin with the ``only if'' direction.  Suppose that $(M,t_a,h^{ab},\nabla,\varphi)$ and $(M,t_a,h^{ab},\nabla',\varphi')$ are both Trautman recoveries of $(M,t_a,h^{ab},\tilde{\nabla})$.  Then by Theorem \ref{recovery}, $\nabla'=(\nabla,C^a{}_{bc})$, where $C^a{}_{bc}=t_bt_c\nabla^a(\varphi'-\varphi)$.  Now suppose that $\xi^a$ is a timelike vector field such that $\nabla^{[a}\xi^{b]}=\mathbf{0}$.  It follows that $\nabla'^{[a}\xi^{b]}=h^{n[a}\nabla'_n\xi^{b]}=-h^{n[a}t_nt_m\nabla^{b]}\xi^m=\mathbf{0}$.  So if $\xi^a$ is non-rotating relative to $\nabla$, it is non-rotating relative to $\nabla'$.  An identical argument establishes the converse, that if $\xi^a$ is non-rotating relative to $\nabla'$, it is non-rotating relative to $\nabla$.

Now consider the ``if'' direction.  Suppose $(M,t_a,h^{ab},\nabla,\varphi)$ is a Trautman recovery of $(M,t_a,h^{ab},\nabla)$ and suppose that $\nabla'$ is a flat derivative operator compatible with $t_a$ and $h^{ab}$ and such that given any timelike vector field $\xi^a$, $\nabla^{[a}\xi^{b]}=\mathbf{0}$ iff $\nabla'^{[a}\xi^{b]}=\mathbf{0}$.  Since $\nabla'$ is compatible with $t_a$ and $h^{ab}$, $\nabla'=(\nabla,C^a{}_{bc})$, where $C^a{}_{bc}=2h^{an}t_{(b}\kappa_{c)n}$ for some smooth antisymmetric tensor field $\kappa_{ab}$ \citep[Prop. 4.1.3]{MalamentGR}.  Now consider some unit timelike vector field $\xi^a$ such that $\nabla^{[a}\xi^{b]}=\mathbf{0}$.  (Such a field always exists because $\nabla$ is flat and $M$ is connected and simply connected.)  It follows that $\mathbf{0}=\nabla'^{[a}\xi^{b]}=\nabla^{[a}\xi^{b]}-2h^{o[b}h^{a]n}t_{(n}\kappa_{m)o}\xi^m=2h^{o[b}h^{a]n}t_{(n}\kappa_{m)o}\xi^m=h^{o[b}h^{a]n}\kappa_{no}\xi^mt_m=\kappa^{ab}$.  This means that $\kappa_{ab}=t_{[a}\sigma_{b]}$, for some covector $\sigma_b$.  (Why?  Because $t_a$ is the \emph{only} covector that annhilates $h^{ab}$.)  It follows that $C^a{}_{bc}=\eta^a t_bt_c$ for some spacelike $\eta^a$ (possibly $\mathbf{0}$).

Now recall that $\nabla$ and $\nabla'$ are both flat by hypothesis, and thus, using the relation $R'^a{}_{bcd}=R^a{}_{bcd}+2\nabla_{[c}C^a{}_{d]b}+2C^n{}_{b[c}C^a{}_{d]n}$ \citep[Eq. 1.8.2]{MalamentGR}, it follows that $2\nabla_{[c}\eta^at_{d]}t_b=\mathbf{0}$.  Acting on both sides with $h^{cn}$, we find $t_b\nabla^{n}\eta^a=\mathbf{0}$.  Since $t_a\neq \mathbf{0}$, it follows that $\nabla^a\eta^b=\mathbf{0}$.  Finally, invoking Prop. 4.1.6 of \citet{MalamentGR}, which holds globally because we have limited attention to spacetimes who spacelike slices are connected and simply connected, we conclude that there exists a smooth scalar field $\tilde{\varphi}$ such that $\eta^a=\nabla^a\tilde{\varphi}$.  Finally, we define $\varphi'=\tilde{\varphi}+\varphi$.  It follows that $\nabla^a\nabla^b(\varphi'-\varphi)=\nabla^a\eta^b=\mathbf{0}$ and that $\nabla'=(\nabla,C^a{}_{bc})$, where $C^a{}_{bc}=t_bt_c\tilde{\nabla}^a(\varphi'-\varphi)$.  Thus, by Theorem \ref{recovery}, $(M,t_a,h^{ab},\nabla',\varphi')$ is a Trautman recovery of $(M,t_a,h^{ab},\tilde{\nabla})$.\hspace{.25in}$\square$

\noindent\textbf{Proof of Prop. \ref{NCrecovery}}

If the curves $\{\gamma\}_{\rho}$ agree with the possible trajectories allowed by ordinary Newtonian gravitation, then there must exist some derivative operator $\nabla\in[\nabla]$ (indeed, by Prop. \ref{MaxwellRecovery}, \emph{any} will suffice) and some $\varphi$, such that $(M,t_a,h^{ab},\nabla,\varphi)$ is a model of Newtonian gravitation.  Prop. \ref{geometrization} then guarantees that there is a unique derivative operator $\nabla$ satisfying (1) and (2) in the proposition.\hspace{.25in}$\square$

\section*{Acknowledgments}
This material is based upon work supported by the National Science Foundation under Grant No. 1331126.  I am grateful to David Malament and Chris Smeenk for helpful conversations related to this paper, and to David Malament and two anonymous referees for detailed comments on a previous draft.

\singlespacing
\bibliography{mhncsk}
\bibliographystyle{elsarticle-harv}

\end{document}